\documentclass{webofc}
\usepackage[varg]{txfonts}   
\usepackage[utf8]{inputenc}  
\usepackage{cancel}
\usepackage{ucs}             
\usepackage[T1]{fontenc}     
\usepackage{lmodern}         
\usepackage{amsmath}         
\usepackage{amssymb}
\usepackage{bbm}             
\usepackage{bm}              
\usepackage[mathscr]{eucal}  
\usepackage{leftidx}         
\usepackage{wrapfig}
\usepackage[usenames]{color} 
\usepackage{comment}         
\usepackage{hyperref}       

\newcommand{\gehtzu}[1]{\stackrel{\footnotesize{#1}}{\to}}

\begin{document}
\title{Calorons, monopoles and stable, charged solitons}



\author{\firstname{Manfried} \lastname{Faber}\inst{1}\fnsep\thanks{\email{faber@kph.tuwien.ac.at}}}

\institute{Technische Universität Wien, Atominstitut, Stadionallee 2, 1020 Wien, Austria}

\abstract{We discuss the similarity of the constituent monopoles of calorons and stable topological solitons with long range Coulombic interaction, classical solutions of the model of topological particles. In the interpretation as electric charges they can be compared to electrons and positrons with spin up and down, with quantised charge and finite mass.}

\maketitle

\section{Caloron monopoles}\label{Sec-Intro}
An important property of SU(2) Yang–Mills theory is the existence of topologically different vacua, characterised by a winding number. There are solutions to the classical equations of motion of this theory in Euclidean space-time, instantons, describing transitions between neighbouring vacua. Instantons have minimal action with the action density concentrated around an event in 4D space-time. They are characterised by a topological quantum number, the topological charge~\cite{BELAVIN197585}. Periodic boundary conditions in Euclidean time model field theories at finite temperatures, where the temperature $T$ is proportional to the inverse time extent. The solutions of the Yang-Mills equations are modified by finite temperature $T$. As shown by Kraan and van Baal~\cite{Kraan:1998pm} and Lee and Lu~\cite{Lee:1998bb} finite $T$ deforms instantons to periodic solutions, calorons. With increasing $T$ calorons separate into constituents, monopoles (dyons), as can be nicely observed in the action density, see e.g. Fig.1 of Ref.~\cite{Kraan:1998pm}. These interesting solutions fire our imagination due to the similarity to localised quantised charges in electrodynamics, atomic and nuclear physics.

On the 4D Euclidean lattice with the gauge field $U_\mu(x)\in SU(2)$ defined on links, calorons are characterised by the Polyakov loops in terms of matrices
\begin{equation}\label{Poly}
Q(\vec x):=\Pi_{i=1}^{N_t}U_4(\vec x,t_i)=q_0(\vec x)-\mathrm i\vec\sigma\vec q(\vec x),\quad q_0^2+\vec q^{\,2}=1,
\end{equation}
by the distribution of these ``unit quaternions'' in 3D space. The constituent monopoles get equal size for $q_0(\infty)=0$. Then their imaginary part $\vec q$ has hedgehog form, as depicted in Table~\ref{TabConst}. There are four topologically different constituents, dubbed $M,\bar M$ and $L,\bar L$~\cite{Larsen2015} according to their electric $e$ and magnetic $m$ charges in SU(2) Yang-Mills theory.
\begin{table}[h!]
\caption{Schemes of the four different topological possibilities of spherically symmetric hedgehogs for unit quaternions approaching $q_0=0$ at infinity. The diagrams show the imaginary components $\vec q$ of the quaternionic field, in full red for the hemisphere with $q_0>0$ and in dashed green for $q_0<0$ and approach black towards $q_0=0$. They differ by parity transformations $\Pi$ and center transformations $z$. The caloron constituents were dubbed $M$ and $L$~\cite{Larsen2015} according to their electric $e$ and magnetic $m$ charges in SU(2) Yang-Mills theory. In the solitonic interpretation they get the charge numbers $Z$ and the topological charge $\mathcal Q$ for electrons and positrons with spin up and spin down.}
\centering
\label{TabConst}
\hspace*{-4.5mm}\begin{tabular}{cccc}\hline 
$\mathcal T=1$&$\mathcal T=\Pi$&$\mathcal T=z$&$\mathcal T=z\Pi$\\\hline\\[-2ex]
$M$&$\bar M$&$L$&$\bar L$\\
$e=1$&$e=1$&$e=-1$&$e=-1$\\
$m=1$&$m=-1$&$m=1$&$m=-1$\\\hline
$\mathcal T=1$&$\mathcal T=z\Pi$&$\mathcal T=z$&$\mathcal T=\Pi$\\\hline
\includegraphics[scale=0.23]{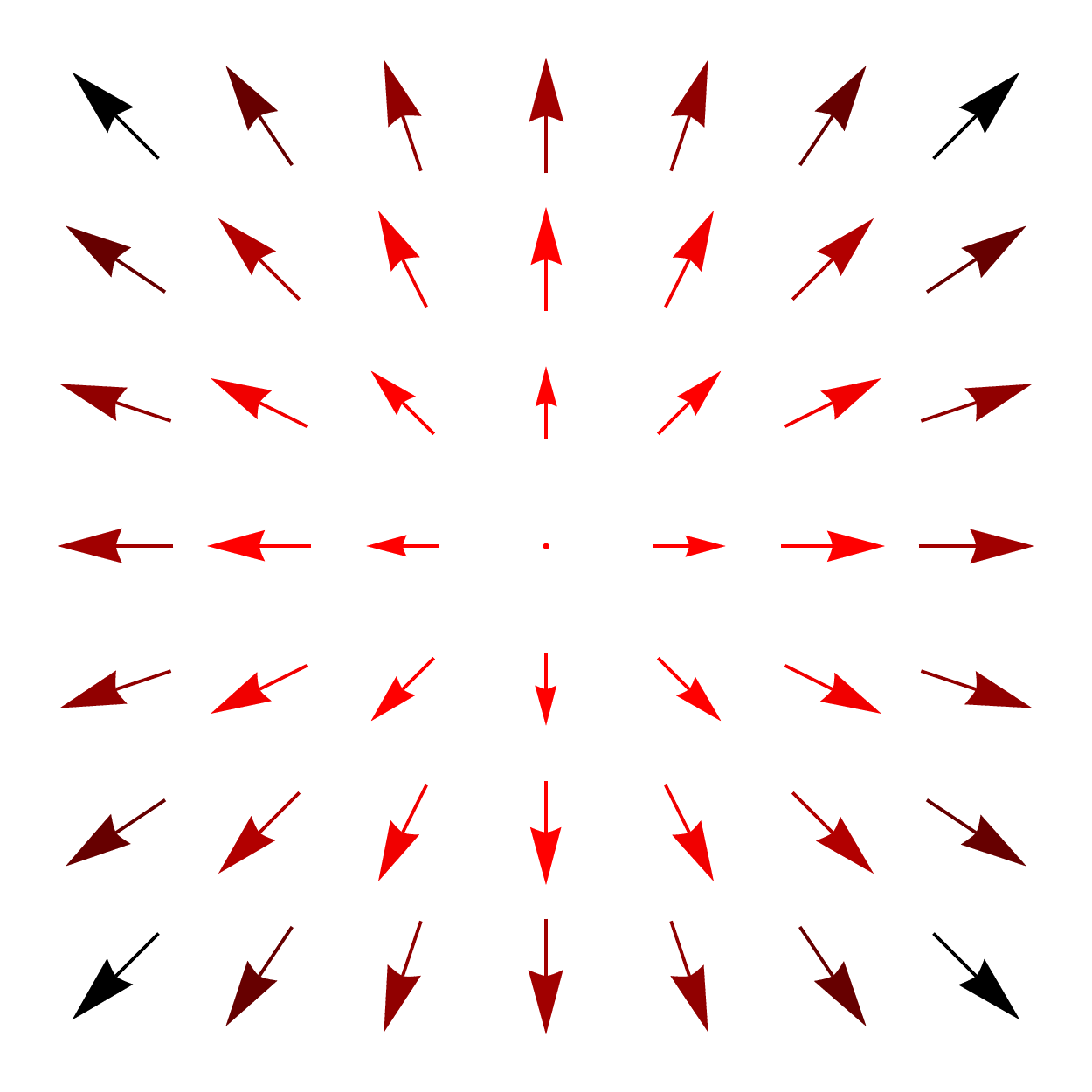}&\includegraphics[scale=0.23]{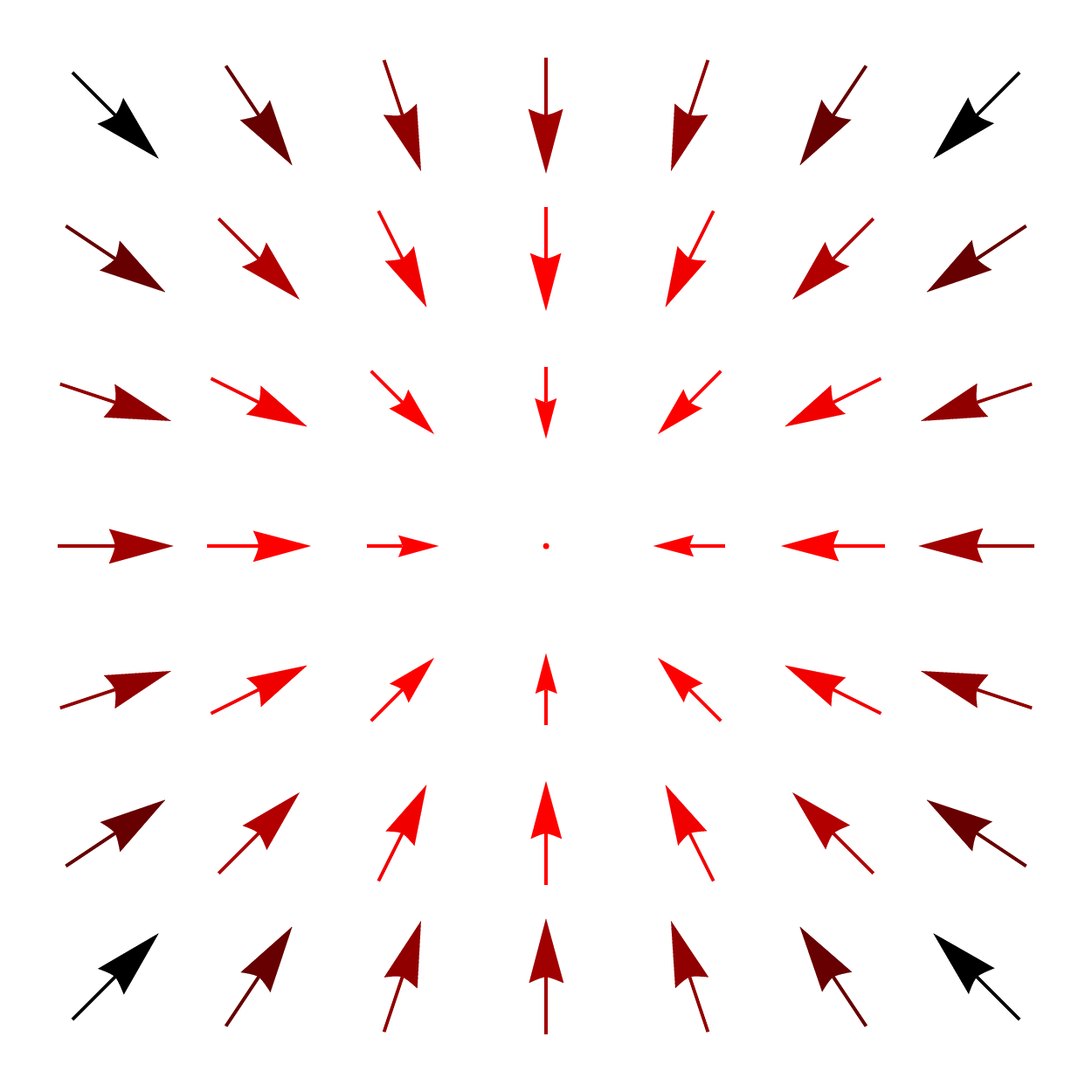}&\includegraphics[scale=0.23]{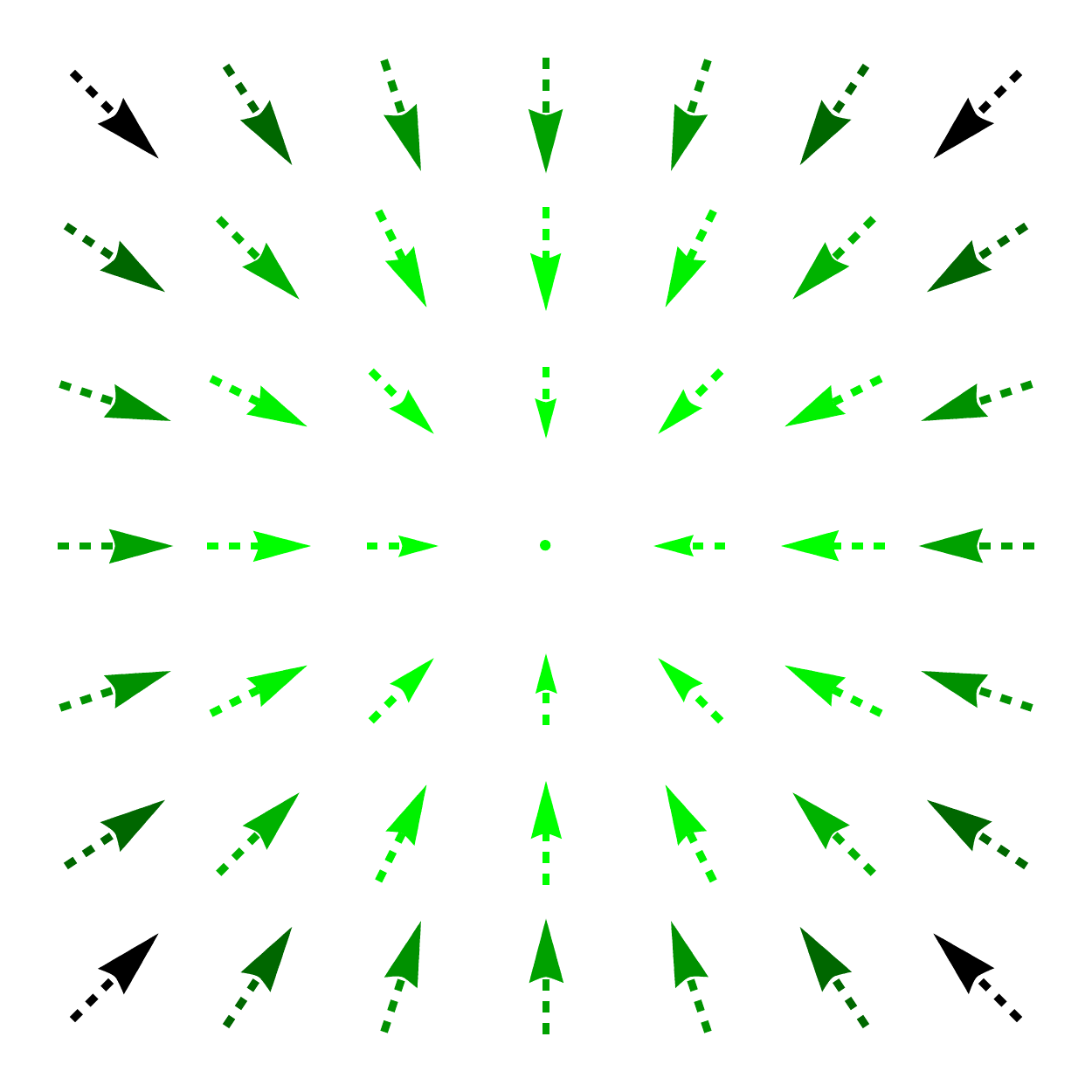}&\includegraphics[scale=0.23]{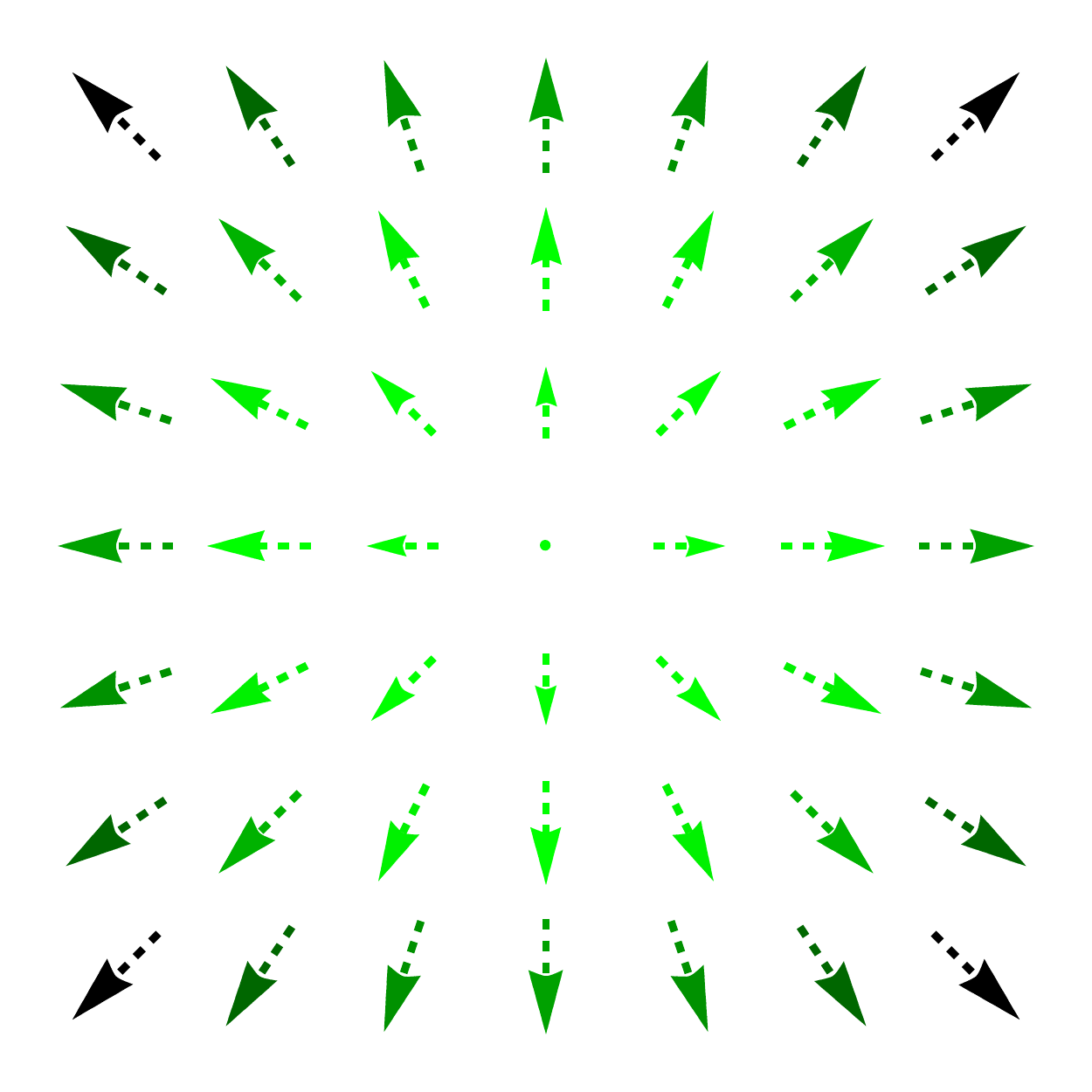}\\\hline
$q_0\ge0$&$q_0\ge0$&$q_0\le0$&$q_0\le0$\\\hline
$Z=1$&$Z=-1$&$Z=-1$&$Z=1$\\
$\mathcal Q=\frac{1}{2}$&$\mathcal Q=-\frac{1}{2}$&$\mathcal Q=\frac{1}{2}$&$\mathcal Q=-\frac{1}{2}$\\\hline
\end{tabular}
\end{table}

In the rest of this article we show that with an appropriate Lagrangian the configurations depicted in Table~\ref{TabConst} can be compared to quantised electric charges, to electrons and postitrons with spin up and down without any divergencies, which QED is plagued with.

\section{Formulation of the model of topological particles (MTP)}\label{Sec-Model}
As described in Ref.~\cite{Faber:2022zwv} MTP uses the SO(3) degrees of freedom (dofs) of spatial Dreibeins to describe electromagnetic phenomena. The calculations are simplified by using SU(2) matrices,
\begin{equation}\label{FeldVariablen}
Q(x)=\mathrm e^{-\mathrm i\alpha(x)\vec\sigma\vec n(x)}
=\underbrace{\cos\alpha(x)}_{q_0(x)}-\mathrm i\vec\sigma\underbrace{\vec n\sin\alpha(x)}_{\vec q(x)}\in SU(2)\cong\mathbb S^3
\end{equation}
in Minkowski space-time as field variables, where arrows indicate vectors in the 3D algebra of su(2) with the basis vectors $\sigma_i$, the Pauli matrices. The Lagrangian of MTP reads,
\begin{equation}
\label{LagrangianMTP}
\mathcal L_{\mathrm{MTP}}(x):=-\frac{\alpha_f\hbar c}{4\pi}
\left(\frac{1}{4}\,\vec R_{\mu\nu}(x)\vec R^{\mu\nu}(x)+\Lambda(x)\right)
\quad\textrm{with}\quad\Lambda(x):=\frac{q_0^6(x)}{r_0^4},
\end{equation}
with the curvature tensor $\vec R_{\mu\nu}$ and the affine connection $\vec\Gamma_\mu$
\begin{equation}\label{curvaturetensor}
\vec R_{\mu\nu}:=\vec{\Gamma}_\mu\times\vec{\Gamma}_\nu\quad\textrm{and}\quad
\left(\partial_\mu Q\right)Q^\dagger=:-\mathrm i\vec\sigma\vec\Gamma_\mu.
\end{equation}
We get contact to nature by defining the electromagnetic field strength tensor by
\begin{equation}\label{FieldStrength}
\vec F_{\mu\nu}:=-\frac{e_0}{4\pi\epsilon_0}^{\star\!\!}\vec R_{\mu\nu}.
\end{equation}
MTP has four different classes of topologically stable single soliton configuration~\cite{Faber:2022zwv}, represented by
\begin{equation}\label{Igel}
n_i(x)=\pm\frac{x^i}{r},\quad\alpha(x)=\frac{\pi}{2}\mp\arctan\frac{r_0}{r}
=\begin{cases}\arctan\frac{r}{r_0}\\\pi-\arctan\frac{r}{r_0}\end{cases}
\end{equation}
The diagrams for the imaginary part of the Q-fields~(\ref{FeldVariablen}) agree with the diagrams for the caloron constituents in Table~\ref{TabConst}\,, but with another interpretation. The four classes differ in two quantum numbers related to charge and spin, as explained in Sect.~\ref{Sec-ChargeSpin}. The configurations within each of the four classes may differ by Poincaré transformations. The rest energy of solitons,
\begin{equation}\label{eRuhenergie}
E_0=\frac{\alpha_f\hbar c}{r_0}\frac{\pi}{4},
\end{equation}
can be adjusted to the electron rest energy $m_ec_0^2=0.511$~MeV by choosing,
\begin{equation}\label{r0m2}
r_0=2.213~\mathrm{fm},
\end{equation}
a scale which is close to the classical electron radius $r_\mathrm{cl}=2.818$~fm. The four parameters, $r_0, c_0, E_0$ and $e_0$, correspond to the natural scales of the four quantities, length, time, mass and charge, of the Système international d’unités, involved in this model. Eq.~(\ref{eRuhenergie}) can therefore be interpreted as a relation between the two fundamental physical quantities $\alpha_f=1/137.036$ and $\hbar$.

How to relate the four classes of solitons to the Dirac equation is discussed in Ref.~\cite{Faber2022Dir}.

\section{Charge and spin}\label{Sec-ChargeSpin}
In the vacuum, at the minimum of the potential, $Q$ is purely imaginary. Therefore, at distances of a few soliton radii $r_0$ the $Q$-field approaches unit vectors $\vec n\in\mathbb S^2$ and the non-abelian field strength tensor $\vec F_{\mu\nu}$ aligns in the direction of $\vec n$, it gets abelian
\begin{equation}\label{asymFeld}
\vec F_{\mu\nu}\gehtzu{r>>r_0}(\vec F_{\mu\nu}\vec n)\vec n.
\end{equation}
This allows us to define the electric current
\begin{equation}\label{EdynGrenJ}
j_q^\kappa:=-\partial_\lambda
\underbrace{\frac{e_0 c}{4\pi}\frac{1}{2}\epsilon^{\kappa\lambda\mu\nu}\,
\vec n[\partial_\mu\vec n\times\partial_\nu\vec n]}_{f^{\kappa\lambda}}
=\frac{1}{\mu_0}\partial_\lambda f^{\lambda\kappa}.
\end{equation}
The sign of the $\vec n$-field~(\ref{Igel}) decides about the charge quantum number $Z$ defined by a map $\Pi_2(\mathbb S^2)$ of closed surfaces around soliton centers to $\mathbb S^2$. For soliton pairs a positive product of their charge numbers leads to repulsion and a negative product to attraction.

Field configurations are further characterised by the number $\mathcal Q$ of coverings of $\mathbb S^3$, by the map $\Pi_3(\mathbb S^3)$,
\begin{equation}\label{EdynGrenJ}
\mathcal Q:=\frac{1}{V(\mathbb S^3)}
\int_0^\infty\mathrm dr\int_0^\pi\mathrm d\vartheta\int_0^{2\pi}\mathrm d\varphi\,
\vec\Gamma_r(\vec\Gamma_\vartheta\times\vec\Gamma_\varphi),
\end{equation}
with values $\pm1/2$ for single solitons configurations~(\ref{Igel}) as listed in Table~\ref{TabConst}. The sign of $\mathcal Q$ defines an internal chirality $\chi:=\mathrm{sign}\mathcal Q$, the chirality of rotations of Dreibeins along the coordinates $x_i$ through the centers of solitons. We interpret the absolute value of $\mathcal Q$ as spin quantum number $s$,
\begin{equation}\label{DefChiral}
\mathcal Q=\chi\cdot s\quad\textrm{with}\quad s:=|\mathcal Q|,
\end{equation}
with the usual addition rules for spins. The spin quantum number of two-soliton configurations indicates that $\chi$ can be related to the sign of the magnetic spin quantum number.

\begin{figure}[h!]
\centering
\includegraphics[scale=0.3]{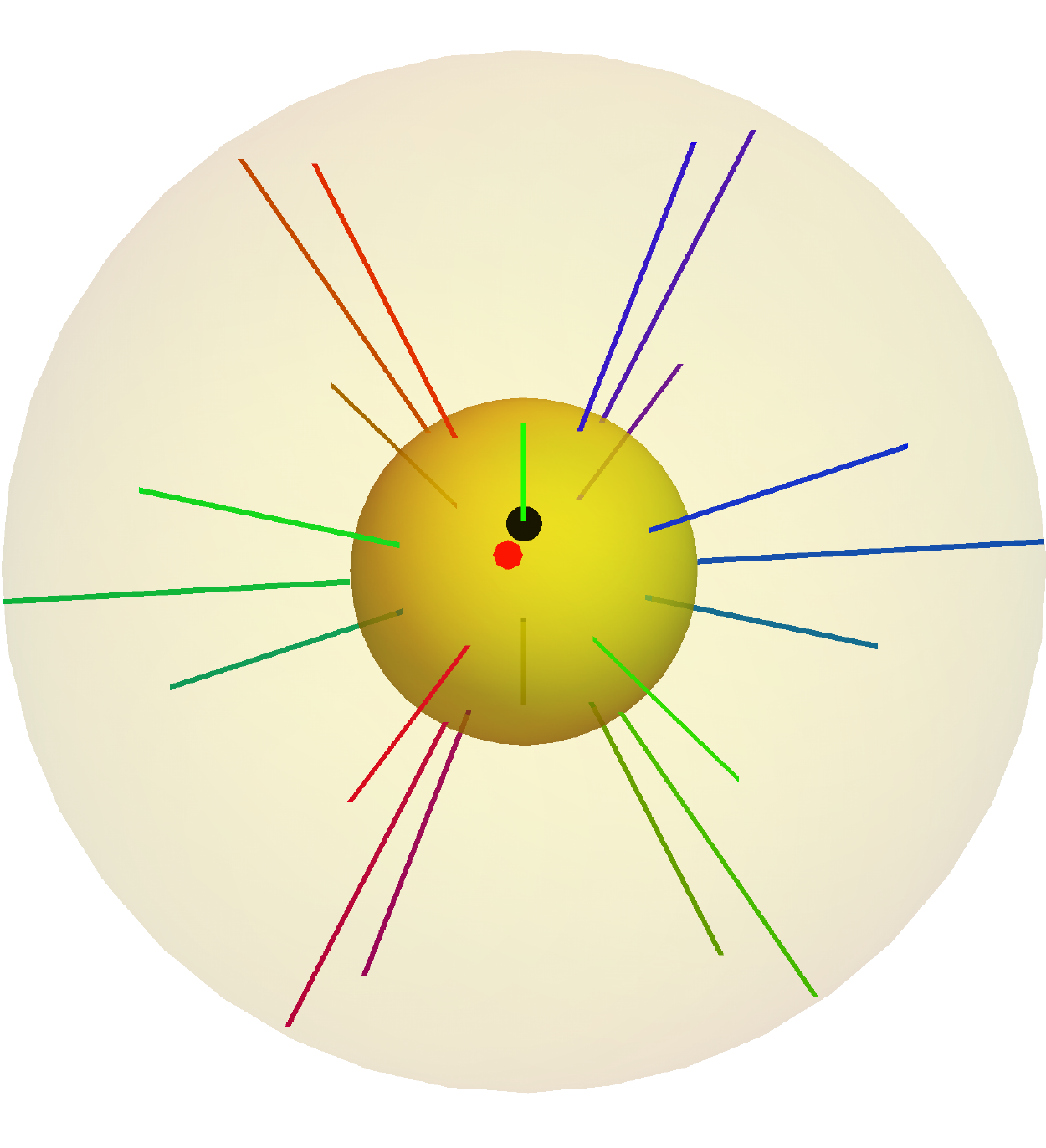}
\includegraphics[scale=0.3]{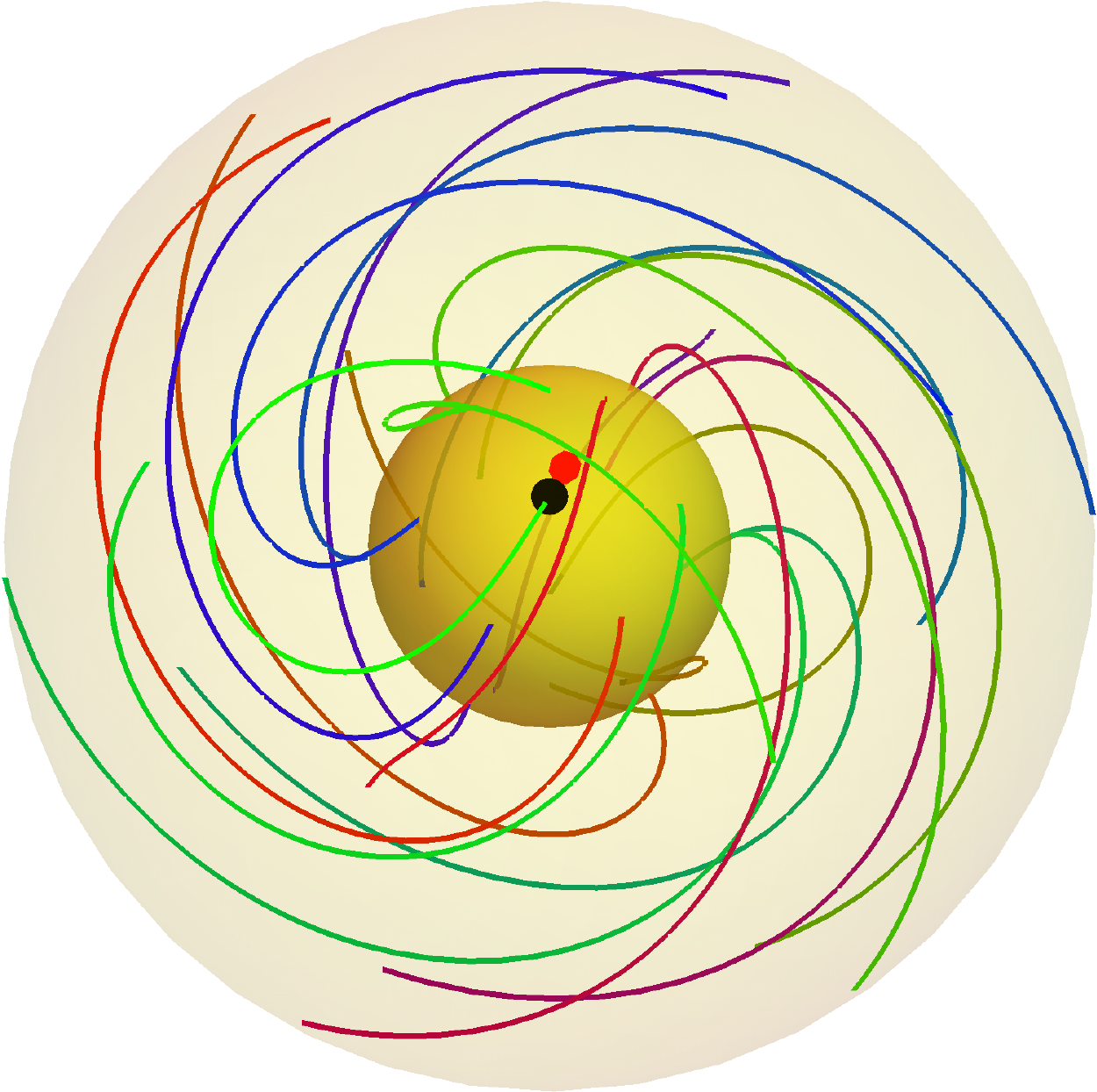}
\includegraphics[scale=0.3]{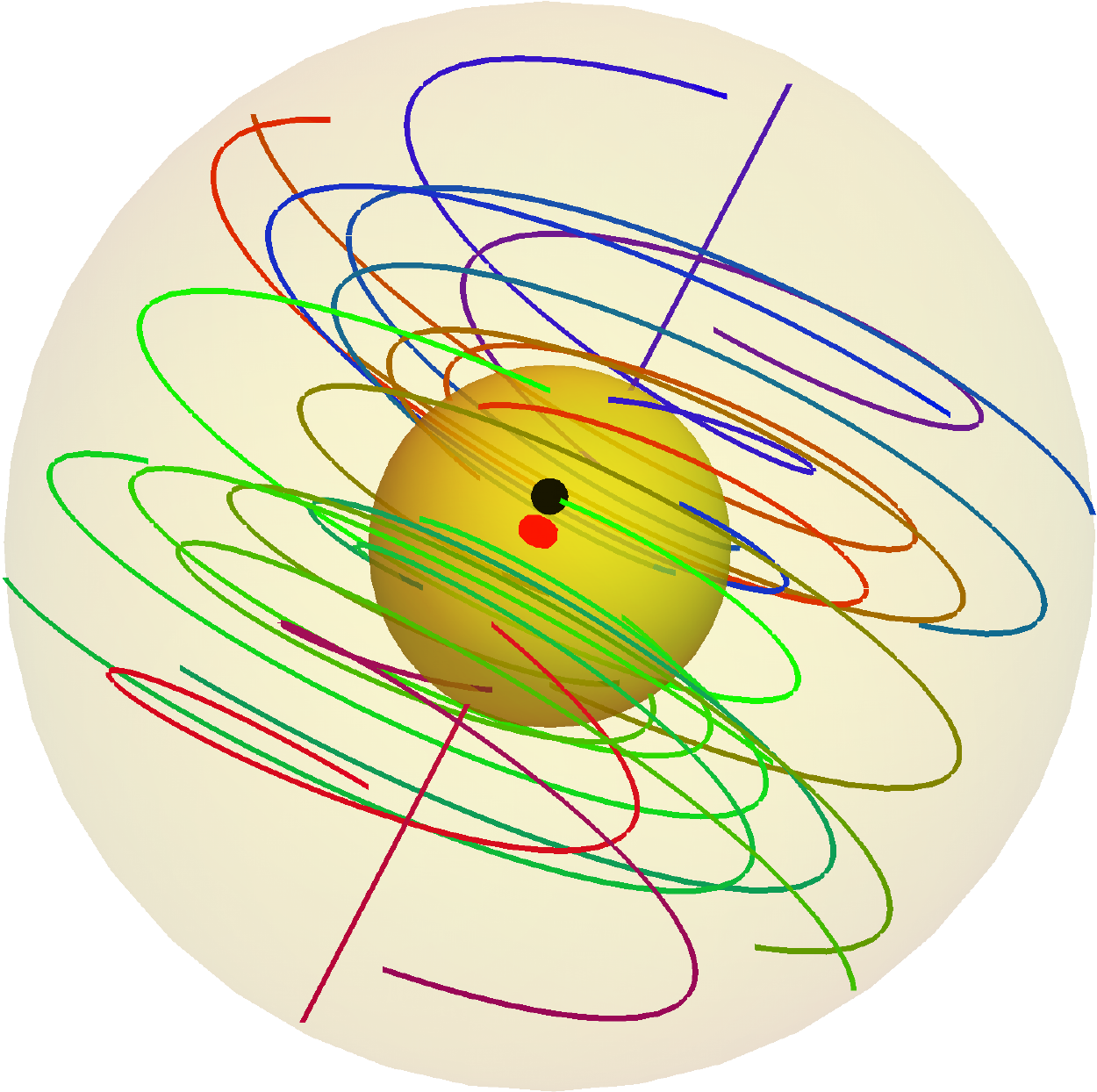}\\
\hfill\includegraphics[scale=0.3]{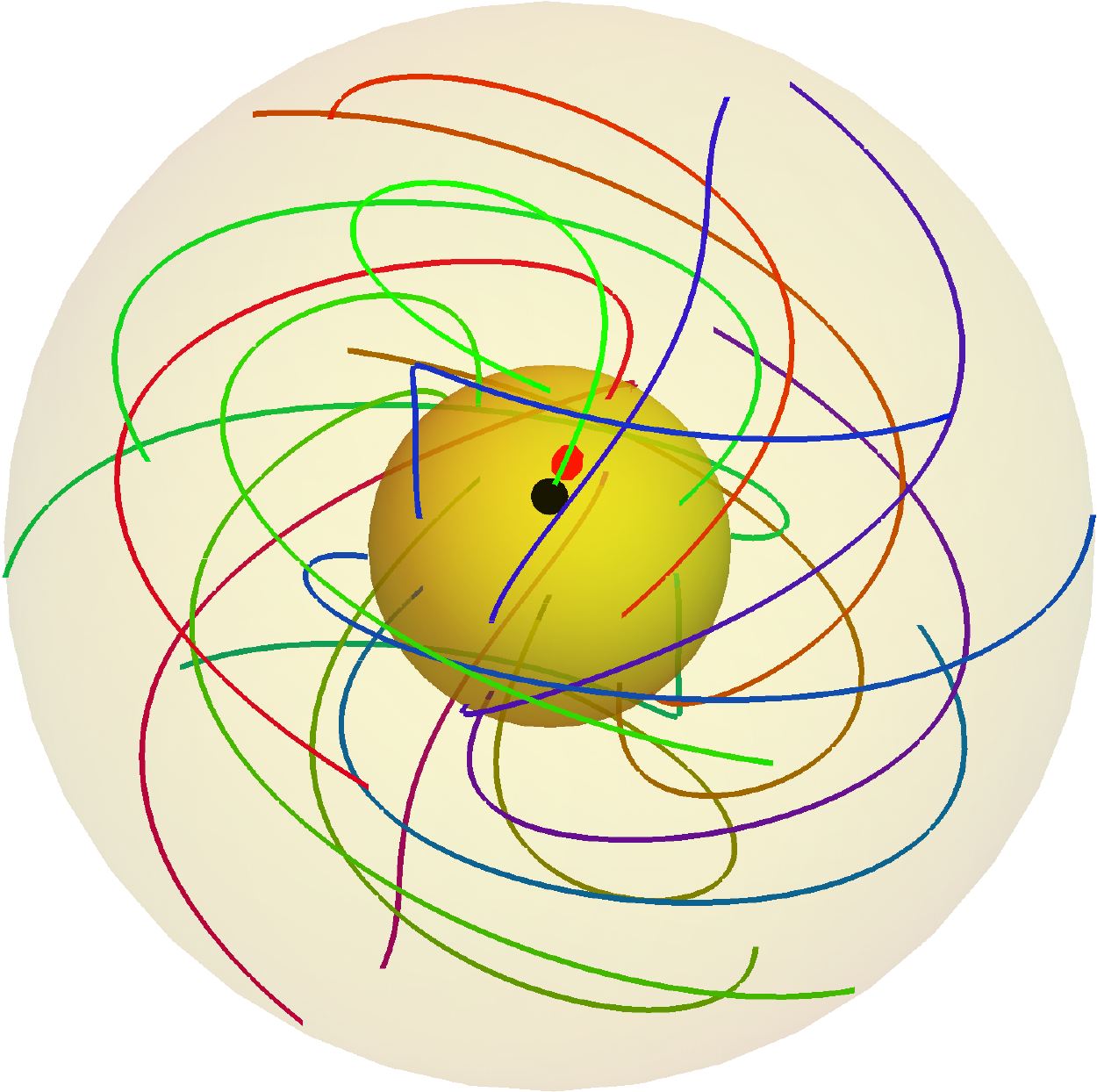}
\includegraphics[scale=0.3]{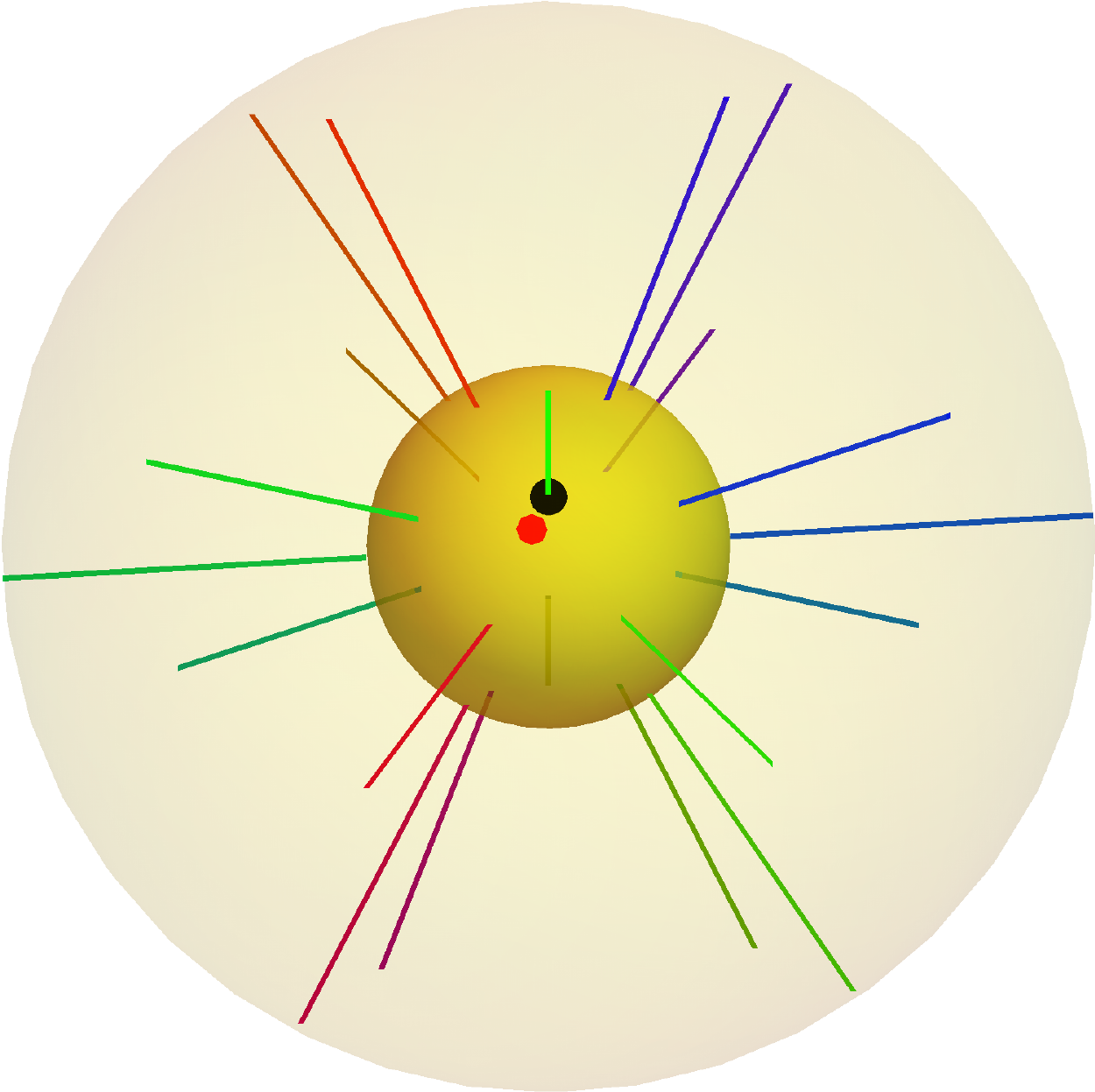}
\caption{The small yellow sphere is connected to the surrounding by wires. If this sphere is rotated as indicated by the red dot, around the axis passing through the black dot, the wires get increasingly entangled up to rotations by $2\pi$ and disentangled again when approaching $4\pi$.}
\label{vierpi}
\end{figure}

Since the solitons of MTP are connected to the surrounding by the lines of constant $\vec n$-field, the interpretation of solitons as spin-1/2 objects is in direct relation to an important property of our 3-dimensional space. It is well known in mathematics that objects, wired to their neighbourhood, can disentangle these wires by their appropriate movement during $4\pi$-rotations, as shown in Fig.~\ref{vierpi}. This property is reflected in the $4\pi$-periodicity of the quantum phase of spin-1/2 particles, of fermions.

\begin{figure}[h!]
\centering
\hspace*{-5mm}\includegraphics[scale=0.50]{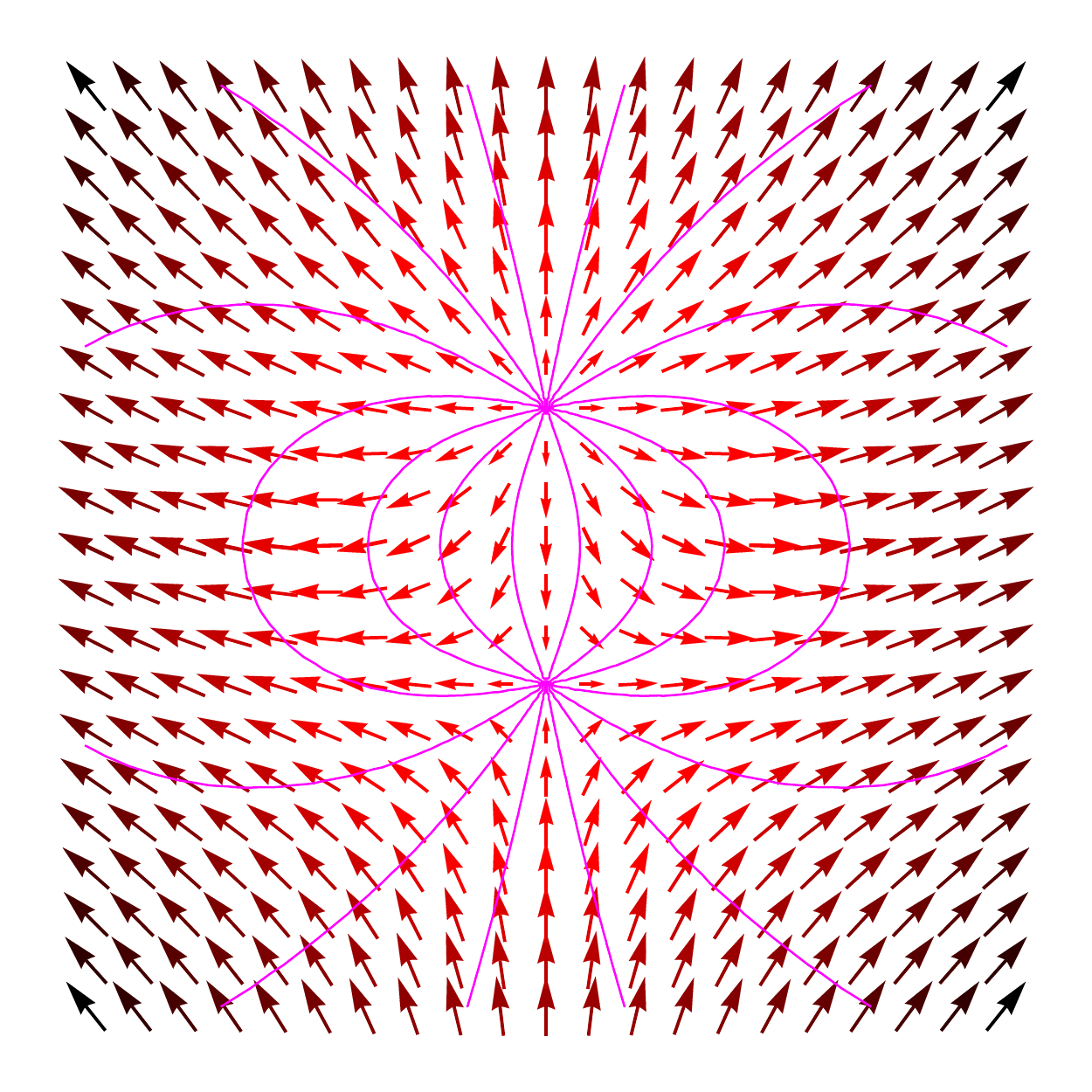}\hspace{5mm}
\includegraphics[scale=0.50]{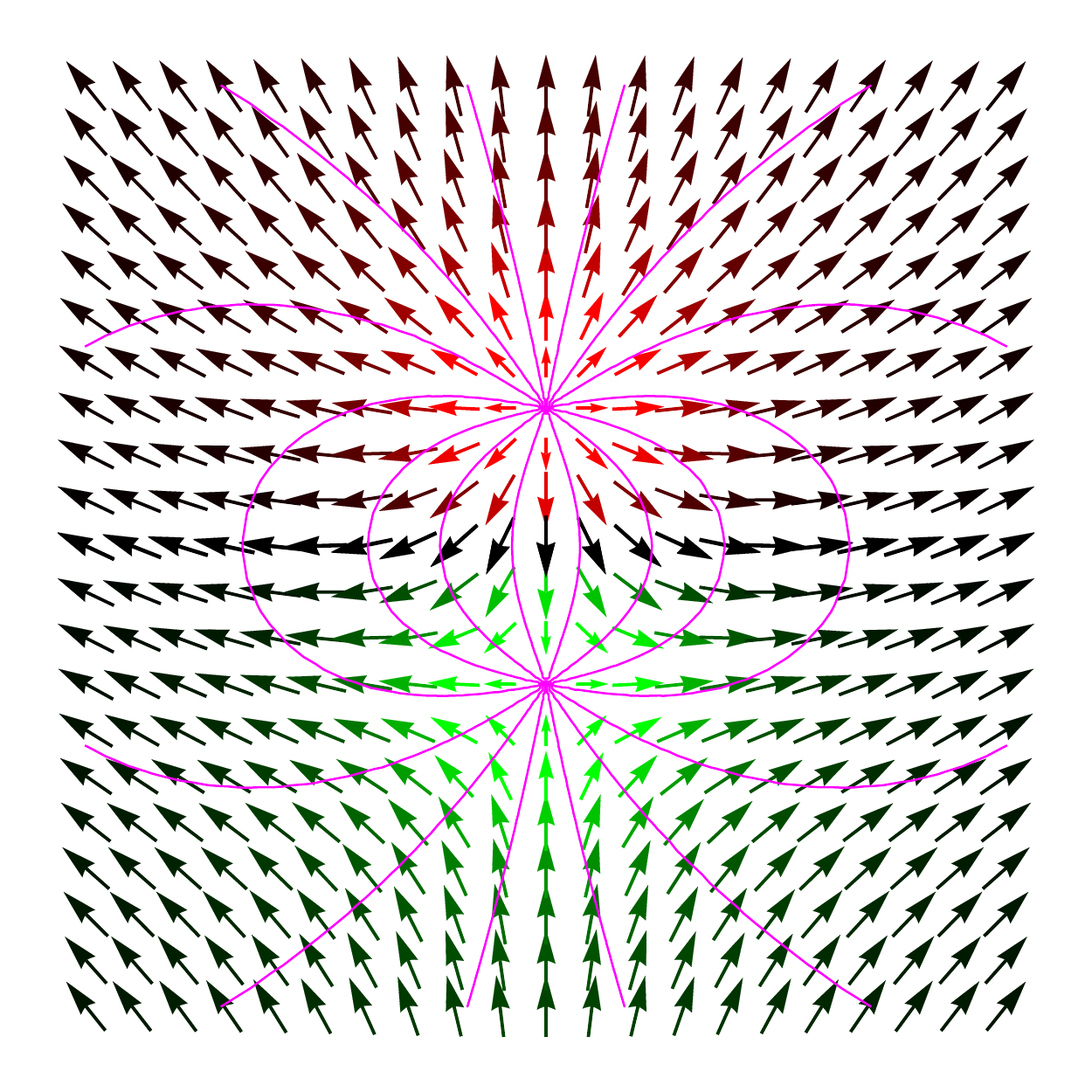}
\caption{Schematic diagrams depicting the imaginary part $\vec q=\vec n\sin\alpha$ of the $Q$-field~(\ref{FeldVariablen}) of two opposite unit charges by arrows. The lines represent some electric flux lines. We observe that they coincide with the lines of constant $\vec n$-field. The configurations are rotational symmetric around the axis through the two charge centres. In the red/green arrows, we encrypt also the positive/negative values of $q_0=\cos\alpha$. For $q_0\to0$ the arrows are getting darker or black. The left configuration belongs to the topological quantum numbers $\mathcal Q=S=0$ and the right one to $\mathcal Q=S=1$, where $S$ is the total spin quantum number of this dipole configuration.}
\label{anziehung}
\end{figure}

To get a feeling of the geometric realisations of charge and spin, we show in Fig.~\ref{anziehung} the schematic diagrams of the $\vec q$-fields of a dipole in the spin-0 and the spin-1 configurations. Due to the structure of the vector field the $S$=0--dipole in the left diagram would fuse in a dynamical calculation since both solitons belong to the $q_0\ge0$ hemisphere of $\mathbb S^3$. The size dependence of the energy of static dipoles of this type and the deviations from the Coulomb law of point-like charges was determined numerically in Ref.~\cite{Wabnig2022}. The spin-1 configuration in the right diagram of Fig.~\ref{anziehung} covers upper and lower hemisphere of $\mathbb S^3$. It is attractive only up to a minimal distance, when repulsion starts due to a drastic increase of the curvature energy.

\begin{figure}[h!]
\centering
\hspace*{-5mm}\includegraphics[scale=0.50]{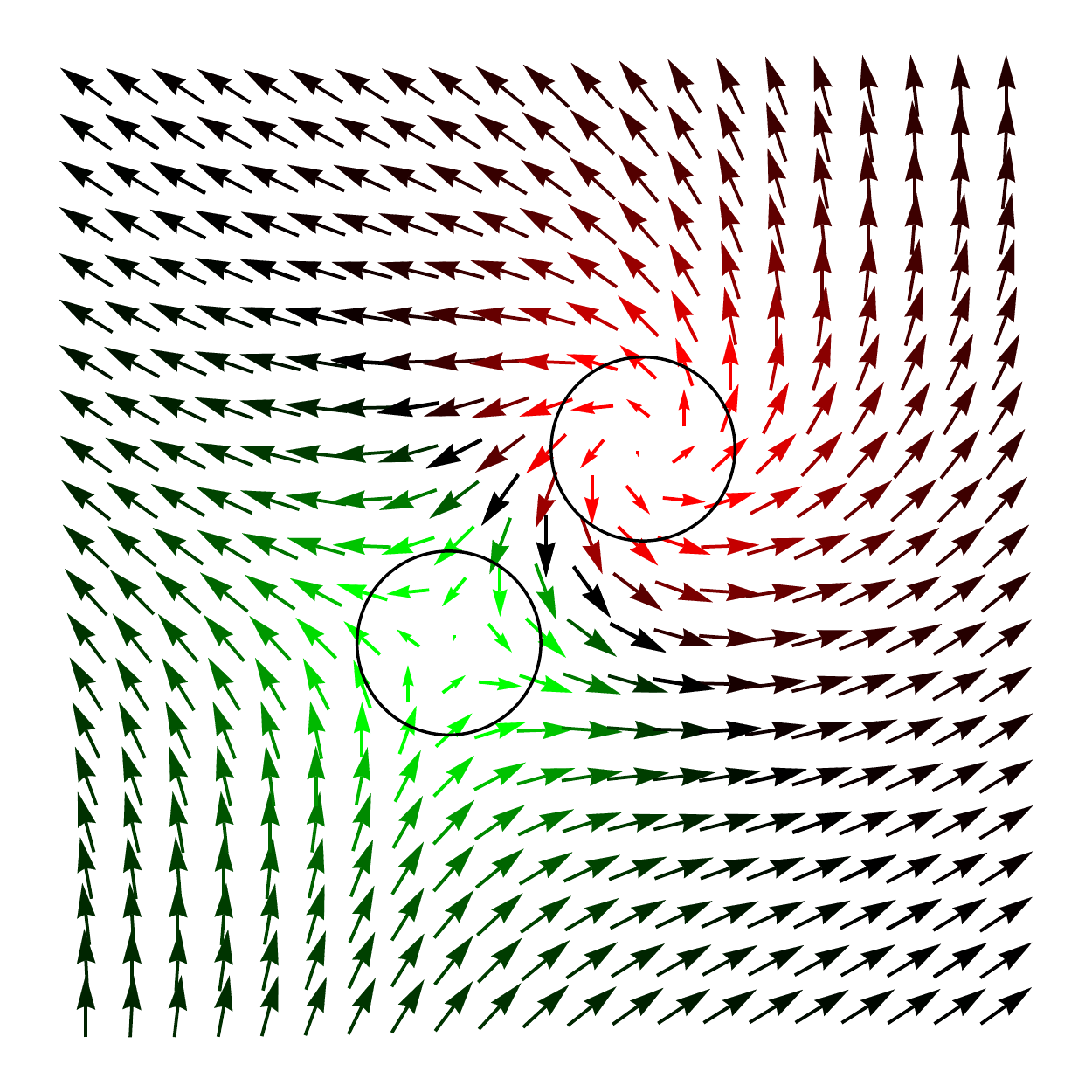}\hspace{5mm}
\includegraphics[scale=0.50]{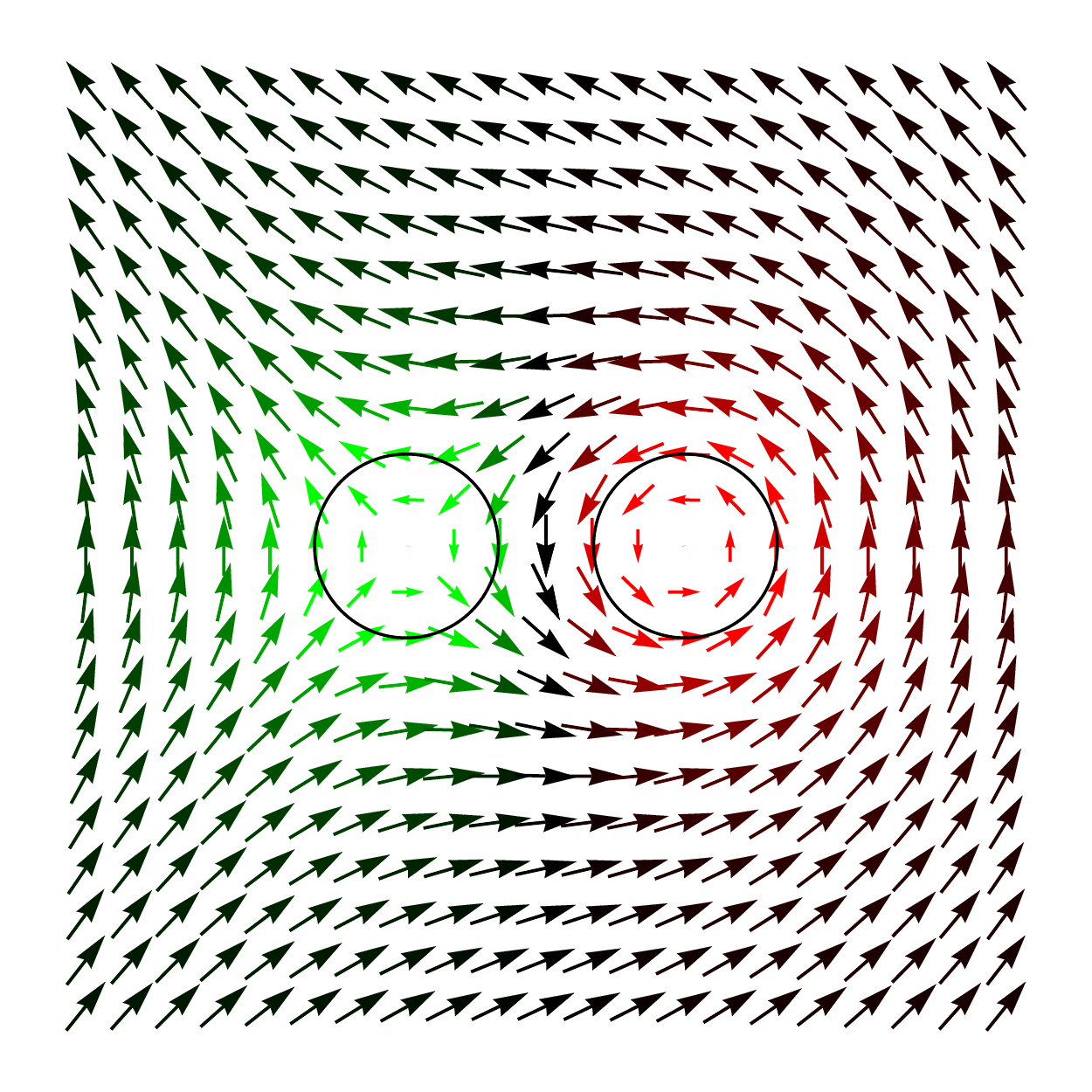}\\
\caption{Schematic diagrams analogous to the right diagram of Fig.~\ref{anziehung} after a 45° and a 90° rotation.}
\label{Rotation}
\end{figure}
According to the definition of the potential term $\Lambda$ in the Lagrangian~(\ref{LagrangianMTP}), MTP has a double degenerate vacuum with two types of Goldstone bosons. The degenerate vacuum has a broken symmetry. In the diagrams in Fig.~\ref{anziehung} this breaking of symmetry is reflected in the direction of the unit vectors at large distances from the soliton centers, they approach asymptotically the vertical direction, $Q(\infty)=-\mathrm i\sigma_3$.

Spin in MTP is a topological quantum number. Furthermore it has to contribute to the total angular momentum. We observe that the static soliton configurations depicted in Table~\ref{TabConst} have no internal rotation and therefore no contribution to the angular momentum. But an orbital angular momentum forces charges to perform an internal rotation due to the broken vacuum which fixes the field at infinity during the rotation. This can be seen, comparing the right diagram of Fig.~\ref{anziehung} with the left diagram of Fig.~\ref{Rotation}, representing a rotation by 45° and the right diagram of Fig.~\ref{Rotation} for a 90° rotation. The internal rotation of the solitons can be observed especially inside the marked circles around the soliton centers. This is a nice picture for a possible realisation of spin in nature.

\section{Comparison to other models}\label{Sec-Vergleich}
MTP with the Lagrangian~(\ref{LagrangianMTP}) is a generalisaton of the Sine-Gordon model~\cite{remoissenet:1999wa}, from 1+1D with one degree of freedom to 3+1D Minkowski space with three SO(3) degrees of freedom (dofs).

With the field variables from SU(2) also the Skyrme model is using the rotational dofs, but with a different Lagrangian. In the Skyrme model~\cite{Skyrme:1958vn,adam:2013tga,gudnason:2022jkn} solitons are compressed by a term quadratic in the field derivatives, the Dirichlet term, which forbids long-range forces. The potential term of MTP allows for Coulombic forces. Both Lagrangians agree in the Skyrme term which tries to smoothen solitons. An essential difference between both models is their vacuum structure. There is only one vacuum in the Skyrme model, whereas there is a two-dimensional manifold of vacua in MTP.

Whether solitons are electric or magnetic is a question of interpretation. The magnetic monopoles which Dirac~\cite{dirac:1931kp,dirac:1948um} introduced are closely related to solitons. Both have quantised charges. But Dirac monopoles have two types of singularities, the Dirac string and the singularity in the center. Dirac strings were removed in two different ways by Wu and Yang. Firstly, by a fiber bundle description~\cite{Wu:1975es} with at least two types of vector fields. In the second method~\cite{Wu:1969wy,Wu:1975vq} they formulated the monopoles with a non-Abelian gauge field. Wu-Yang monopoles still suffer from the singularity in the center. The solitons of MTP are soft-core Dirac monopoles, where all singularities are removed.

Polyakov~\cite{Polyakov:1974ek} and 't Hooft~\cite{tHooft:1974qc} identified monopoles in the Georgi-Glashow model. This model has 15 dofs, a triplet of gauge fields and a triplet of scalar (Higgs) fields. The mass of these monopoles is of the order of the mass of the W-boson multiplied by the inverse of Sommerfeld's fine structure constant~\cite{Polyakov:1974ek}. The interaction between the monopoles is a function of the charge and of the properties of the Higgs field.

The basic fields of MTP are the rotational degrees of freedom of  spatial Dreibeins. Therefore, both long-range forces, gravitation and electromagnetism, are formulated with the properties of space-time only.

The immense success of Maxwell's electrodynamics requires a comparison with MTP. Due to the lack of space we can only enumerate agreements and differences. More details are exposed in Ref.~\cite{Faber:2022zwv}.

\subsubsection*{In agreement with Maxwell's electrodynamics}\label{Sec-EMVergleich}
\begin{itemize}
\item The Lagrangian is Lorentz covariant, thus the laws of special relativity are respected.
\item Charges have Coulombic fields fulfilling Gauß' law.
\item Charges interact via $O(\frac{1}{r^2})$ electric fields, they react to Coulomb and Lorentz forces.
\item A local U(1) gauge invariance is respected.
\item There are two dofs of massless excitations for photons.
\end{itemize}

\subsubsection*{Differences to Maxwell's electrodynamics}\label{Sec-Unterschiede}
\begin{itemize}
\item Field energy is the only source of mass.
\item Charges and their fields are described by the same dofs. Therefore charges cannot be separated from their fields.
\item Charges are quantised in analogy to Dirac monopoles.
\item The self-energy of elementary charges is finite and does not need regularisation and renormalisation.
\item The finite size of solitons leads to a running of the charge.
\item Field dofs can be interpreted as orientations of spatial Dreibeins.
\item The mirror properties of  particles and antiparticles~\cite{Borchert2022} are explained by their topological construction.
\item Particles are characterised by topological quantum numbers.
\item Solitons and antisolitons have opposite internal parity.
\item Spin has usual quantisation properties and  combination rules.
\item Spin contributes to angular momentum due to internal rotations.
\item Solitons are characterised by a chirality quantum number which can be related to the sign of the magnetic quantum number.
\item The quantum numbers of the four classes of soliton configurations agree with the quantum numbers of fields in Dirac spinors.
\item The canonical energy-momentum tensor is automatically symmetric.
\item Static charges are described by the spatial components of vector fields.
  Moving charges need time-dependent fields.
\item Local U(1) gauge invariance emerges from choice of bases on $\mathbb S^2$.
\item Photon number is given by the Gauß\-ian linking number of fibres on $\mathbb S^2$.
\item Photon number changes by interaction with charges.
\item Spin and magnetic moment are dynamical properties only.
\item Electric and magnetic field vectors are perpendicular to each other.
\item Existence of unquantised magnetic currents is allowed.
\item $\alpha$-waves in $q_0=\cos\alpha$ could contribute to the (dark) matter density.
\item $\alpha$-waves lead to additional forces acting on particles and are a possible origin of quantum fluctuations.
\item Potential term suggests a mechanism of cosmic inflation.
\item Potential term contributes to the dark energy.
\end{itemize}

\section{Conclusion}\label{Sec-Zusammenfassung}

A fascination of calorons and their constituent monopoles is due their similarity to the properties which we expect to hold for elementary charges, aspects which are not reflected in our present picture of elementary particle physics. These features originate in the SU(2) manifold and the topological nature of calorons.

Calorons have these interesting properties in common with the model of topological particles (MTP), similarly formulated with SO(3) or its double covering, SU(2). This article gives a short introduction to MTP and enumerates its properties. Some of these properties we expect to hold for elementary charges, some seem to disagree at first look and some are rather speculative. Being optimistic one can hope that MTP is able to improve on our understanding of basic properties of nature by its geometric properties and fills some gaps, left in our present rather algebraic description of nature.

Possibly Galileo is more right than we thought, when he wrote in Il Saggiatore: "Philosophy is written in this grand book, the universe \dots\; It is written in the language of mathematics, and its characters are triangles, circles, and other geometric figures \dots"\cite{discoveries1957}.

\bibliography{caloronsSolitons}
\end{document}